\documentclass[a4paper,preprintnumbers,floatfix,superscriptaddress,pra,twocolumn,showpacs]{revtex4-1}

\usepackage{amsmath, amsthm, amssymb}
\usepackage{graphicx}
\usepackage{dcolumn}
\usepackage{bm}
\usepackage{color}
\usepackage[usenames,dvipsnames]{xcolor}
\usepackage{hyperref}

\newcommand{\eqnref}[1]{(\ref{#1})}
\newcommand{\figref}[1]{Fig.~\ref{#1}}

\newcommand{\ket}[1]{| #1 \rangle}

\newcommand{\bra}[1]{\langle #1 |}

\renewcommand{\t}[1]{\textrm{#1}}

\DeclareGraphicsExtensions{.png}

\begin{document}
\title{Noisy metrology beyond the standard quantum limit}

\author{R. Chaves}\affiliation{ICFO-Institut de Ci\`encies Fot\`oniques, Mediterranean Technology Park, 08860 Castelldefels (Barcelona), Spain}\affiliation{Institute for Physics, University of Freiburg, Rheinstrasse 10, D-79104 Freiburg, Germany}
\author{J. B. Brask}\affiliation{ICFO-Institut de Ci\`encies Fot\`oniques, Mediterranean Technology Park, 08860 Castelldefels (Barcelona), Spain}
\author{M. Markiewicz}\affiliation{Institute of Theoretical Physics and Astrophysics, University of Gda\'nsk, 80-952 Gda\'nsk, Poland}
\author{J. Ko\l{}ody\'{n}ski}\affiliation{Faculty of Physics, University of Warsaw, 00-681 Warszawa, Poland}
\author{A. Ac\'in}\affiliation{ICFO-Institut de Ci\`encies Fot\`oniques, Mediterranean Technology Park, 08860 Castelldefels (Barcelona), Spain}
\affiliation{ICREA-Instituci\'o Catalana de Recerca i Estudis Avan\c cats, Lluis Companys 23, 08010 Barcelona, Spain}

\begin{abstract}
Parameter estimation is of fundamental importance in areas from atomic spectroscopy and atomic clocks to gravitational wave-detection. Entangled probes provide a significant precision gain over classical strategies in the absence of noise. However, recent results seem to indicate that any small amount of realistic noise restricts the advantage of quantum strategies to an improvement by at most a multiplicative constant. Here we identify a relevant scenario in which one can overcome this restriction and attain super-classical precision scaling even in the presence of uncorrelated noise. We show that precision can be significantly enhanced when the noise is concentrated along some spatial direction, while the Hamiltonian governing the evolution which depends on the parameter to be estimated can be engineered to point along a different direction. In the case of perpendicular orientation, we find super-classical scaling and identify a state which achieves the optimum.
\end{abstract}

\maketitle

%Introduction

\begin{figure} [h!]
\centering
\includegraphics[width=1.0\linewidth]{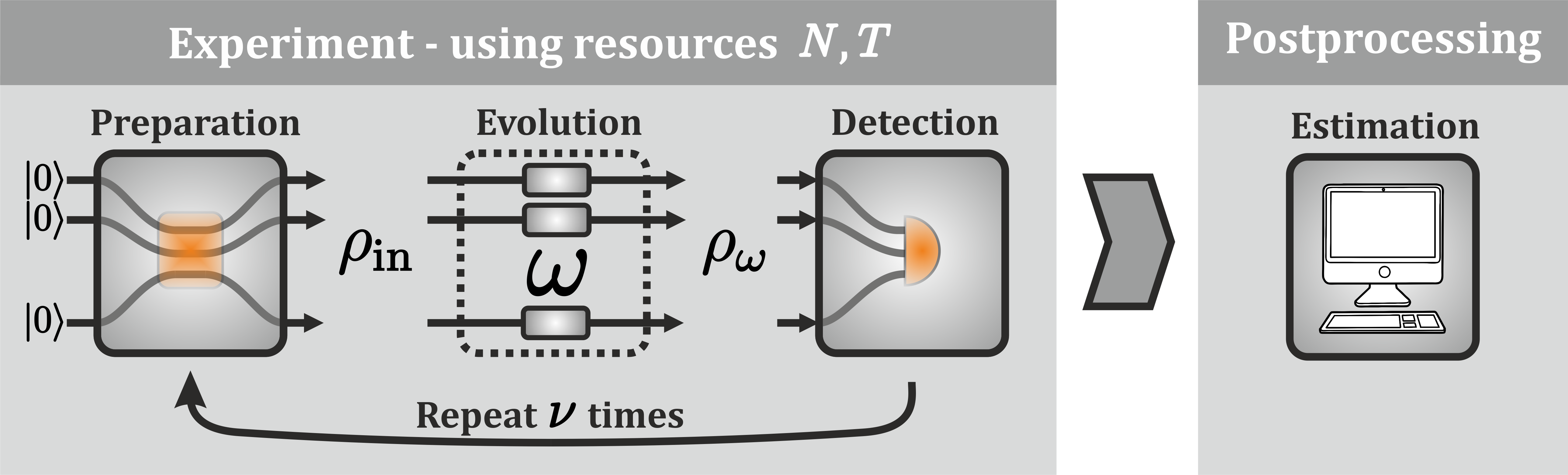}
\caption {General metrology protocol where a known probe state evolves according to a physical evolution depending on an unknown parameter $\omega$. After sufficient amount of data is collected an estimate for the parameter is constructed.} \label{fig:metrology}
\end{figure}

Estimation of an unknown parameter is essential across disciplines from atomic spectroscopy and clocks~\cite{huelga1997, bollinger1996, buzek1999} to gravitational wave-detection~\cite{LIGO2011}. It is typically achieved by letting a probe, e.g.~light, interact with the system under investigation, picking up information about the desired parameter. As seen in Fig. \ref{fig:metrology}, a metrology protocol can be understood in four main steps ~\cite{giovannetti2006,giovanetti2011}: i) preparation of the probe, ii) interaction with the system, iii) readout of the probe, and iv) construction of an estimate of the unknown parameter from the results. Steps (i)-(iii) may be repeated many times before the final construction of the estimate.

The estimate uncertainty will depend on the available resources, here the probe size $N$ and the total time $T$ available for the experiment (other choices are possible~\cite{shaji2007}). By the central limit theorem, for $N$ uncorrelated particles, the best uncertainty scales as $1/\sqrt{\nu N}$, where $\nu=T/t$ is the number of evolve-and-measure rounds. This bound is known as the shot-noise or standard quantum limit (SQL). By making use of quantum phenomena, a metrology protocol may surpass the SQL, reaching instead the limits imposed by the quantum uncertainty relations. For probes of non-interacting particles, the best possible scaling compatible with these relations is $1/(\!\sqrt{\nu}N)$, known as the Heisenberg limit.

Without noise, the Heisenberg limit can be attained using entangled input states, e.g.~Greenberger-Horne-Zeilinger (GHZ) states for atomic spectroscopy~\cite{GHZ}. In the presence of noise however, the picture is much less clear, as the optimal strategy depends strongly on the model of decoherence considered. Nevertheless, the SQL has been significantly surpassed in  experiments of optical magnetometry~\cite{wolfgramm2010,wasilewski2010}, which proved that some sources of noise can be effectively counterbalanced~\cite{andre2004,auzinsch2004}. However, unless one can keep improving the interaction strength or readout efficiency with probe size, e.g.~increasing the optical depth with the atom number in atomic vapours, destructive effects of uncorrelated noise are bound to dominate at high $N$. In this regime of fixed noise (independent of the particle number), a number of no-go results exist, demonstrating that for most types of noisy channels acting independently on each probe particle, an infinitesimally small amount of decoherence limits any quantum improvement over the SQL to at most a constant factor ~\cite{huelga1997,kolodynski2010,*knysh2011, escher2011,demkowicz2012,fujiwara2008,ji2008,hayashi2011}. In particular, any noisy evolution described by $N$ full rank channels, i.e.~channels for which no subspace of the probe state space is free of decoherence, belongs to this class~\cite{demkowicz2012,fujiwara2008,ji2008,hayashi2011}. This is arguably the most likely evolution in experiments, suggesting that the precision is always bound to scale classically for large enough $N$.

Here, for a frequency estimation task undergoing uncorrelated noise and for a configuration where the Hamiltonian and the noise have preferred spatial directions transversal to each other, we show that the restriction to SQL-like scaling can be surpassed. This is achieved by optimising the duration $t$ of the evolve-and-measure rounds. As $N$ increases, the quantum channel describing the single-particle evolution varies due to this optimization. This allows circumventing the conditions of previous no-go results which were derived assuming a \emph{fixed} form of the $N$ single-particle channels \cite{demkowicz2012,fujiwara2008,ji2008,hayashi2011}. Although $t$-optimisation has been considered previously and was not sufficient on its own \cite{huelga1997,escher2011}, we demonstrate that in combination with directionality of the noise it enables beating the SQL. The corresponding channel is full rank, yet we find that a GHZ-state input attains a precision scaling asymptotically as $1/N^{5/6}$. This is found numerically and confirmed by a semi-analytical argument. We further demonstrate numerically that this scaling is optimal by identifying an upper bound on the precision which is saturated by the GHZ state. We also analyse deviations from perfectly directional noise. The asymptotic scaling is then again restricted to SQL-like. However, for small deviations, we observe a much higher precision gain than for parallel noise. Note that here the noise is purely Markovian. Exploting non-Markovianity can also lead to improved precision scaling, as previously shown \cite{matsuzaki2011,*chin2012}.

%Model and Methods

\textit{Model for noisy frequency estimation---}
We consider a Hamiltonian $H=\frac{\omega}{2} {\sum_{k=1}^{N}} \sigma_{z}^{k}$, where $\sigma_z^k$ is a Pauli operator acting on the $k$'th spin-1/2 particle (qubit), and $\omega$ is an unknown frequency to be estimated. To account for noise, we model the evolution by a master equation of Lindblad form
\begin{equation}
\label{eq.master}
\frac{\partial \rho \left( t\right) }{\partial t}=\mathcal{H}\left( \rho \right) +\mathcal{L}\left( \rho \right) .
\end{equation}
Here, $\mathcal{H}\left( \rho\right)= -i\left[ H,\rho \right]$ describes unitary evolution and the Liouvillian $\mathcal{L}(\rho)$ describes noise. We consider uncorrelated noise, such that $\mathcal{L}= {\sum_{k}} \mathcal{L}^{k}$ and for a single qubit we have
\begin{equation}
\mathcal{L}^{k}\rho =-\frac{\gamma }{2}\left[ \rho - \alpha_{x} \sigma^{k}_{x}\rho \sigma^{k} _{x}- \alpha_{y} \sigma^{k} _{y}\rho \sigma^{k} _{y} - \alpha_{z} \sigma^{k}_{z}\rho \sigma^{k} _{z} \right],
\label{lindblad}
\end{equation}
where $\gamma$ is the overall noise strength and $\alpha_{x,y,z} \geq 0$ with $\alpha_{x}\!+\!\alpha_{y}\!+\!\alpha_{z}\!=\!1$. For $\alpha_{z}\!=\!1$ this describes the situation considered by Huelga \textit{et al.}~\cite{huelga1997}, namely dephasing along the direction of the unitary, while $\alpha_{x}\!=\!1$ corresponds to dephasing transversal to the unitary. The latter model resembles the magnetometry setup of \cite{wasilewski2010}, in which the estimated magnetic field is directed perpendicularly to the dominant dephasing dictating the spin decoherence time and the spin relaxation is ignored. For $\alpha_{x}\!=\!\alpha_{y}\!=\!\alpha_{z}\!=\!1/3$ we have an isotropic depolarizing channel.

The uncertainty $\delta\omega$ in the estimate of $\omega$ can be expressed in terms of the quantum Fisher information~\cite{holevo1982} (QFI) $\mathcal{F}(\rho_\omega)$ of the probe state $\rho_\omega$ after evolution. According to the quantum Cram\'er-Rao inequality~\cite{braunstein1994}
\begin{equation}
\label{eq.qcrbound}
\delta\omega \, \sqrt{T} \geq \frac{1}{\sqrt{\mathcal{F}(\rho_\omega) / t}} ,
\end{equation}
a bound achievable asymptotically for $\nu \gg 1$~\cite{braunstein1996,paris2009}. One can explicitly solve the master equation \eqnref{eq.master}, obtaining a map for the corresponding channel (see Appendix A for details). Applying the map to a given input, one obtains $\rho_{\omega}$, and the QFI is computed through the diagonalization of this state. Alternatively, bounds on the QFI can be computed from the Kraus representation of the channel~\cite{escher2011,demkowicz2012}. We write $\mathcal{F}_N = \mathcal{F}(\rho_\omega)$ for probes of $N$ particles.

\begin{figure}
\centering
\includegraphics[width=0.75\linewidth]{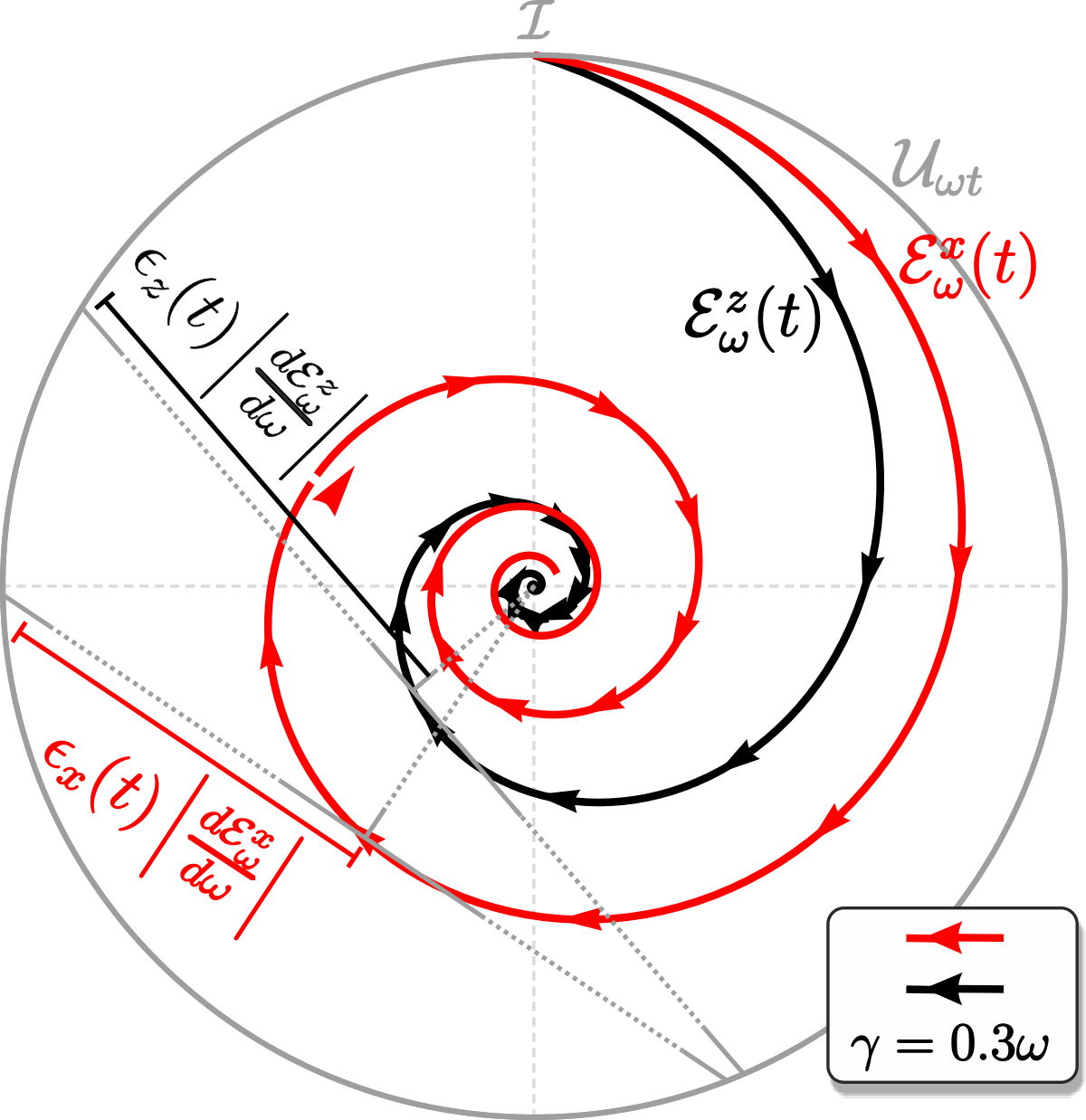}
\caption{Cut through the space of all valid qubit maps. The identity channel $\mathcal{I}[\varrho]\!=\!\varrho$ is at the top, and (extremal) unitary channels $\mathcal{U}_{\omega t}$ are at the circumference. One can think of phase $\omega t$ as increasing along the angular direction and decoherence $\gamma t$ along the radial direction. The spirals represent trajectories of channels for parallel $\mathcal{E}_{\omega}^{z}(t)$ and transversal $\mathcal{E}_{\omega}^{x}(t)$ noise, approaching the completely mixing channel at the disk centre. Arrows are separated by fixed time steps. We see that $\mathcal{E}_{\omega}^{x}(t)$ loses coherence slower, being always closer to the boundary, and also has larger angular `speed' $\left|{d\mathcal{E}_{\omega}^{x}}/{d\omega}\right|$. The classical simulation method~\cite{demkowicz2012} provides a geometric bound on estimation precision based on the distance to the boundary: $\delta\omega\sqrt{T}\!\ge\!\!\sqrt{(\epsilon(t)^{2}t)/N}$. For asymptotic $N$, the bound is determined by the behaviour for small $t$ at the start of the spirals. We get $\epsilon_{z}(t)^{2}\!=\!{2\gamma}/{t}\!+\! O(1)$ and $\epsilon_{x}(t)^{2}\!=\!(\gamma^{2}\omega^2t^{2})/{12}\!+\! O(t^{4})$.}
\label{fig.CS}
\end{figure}

For a general input state, it is a difficult task to compute the QFI, as the size of $\rho_\omega$ grows exponentially with $N$. However, for a GHZ state,
\begin{equation}
\label{ghzstate}
\ket{GHZ} = \frac{1}{\sqrt{2}}\left(\ket{0}^{\otimes N} +\ket{1}^{\otimes N} \right)
\end{equation}
the calculation simplifies dramatically (see Appendix B), and we are then able to optimise $\mathcal{F}_N/t$ (and hence $\delta\omega$) for large $N$. To obtain results valid for general inputs, we resort to bounds on the precision. A first indication that performance is better under transversal than parallel noise is given by the geometric classical simulation (CS) method~\cite{demkowicz2012} (see \figref{fig.CS}). We obtain tighter bounds by adapting the finite-$N$ channel extension (CE) method of Kolodynski \textit{et al.}~\cite{kolodynski2012} to allow for optimization over $t$ (see Appendix C). As for previous asymptotic methods~\cite{escher2011,demkowicz2012} without $t$-optimization, we cannot apriori guarantee our bound to be saturable. However, for parallel noise ($\alpha_z=1$), our bound is known to be tight in the asymptotic $N$ limit both with~\cite{ulam2001} and without~\cite{kolodynski2012} $t$-optimisation. As we find below, the bound is also tight for transversal noise.

\begin{figure*} [!t]
\centering
\includegraphics[width=0.95\textwidth]{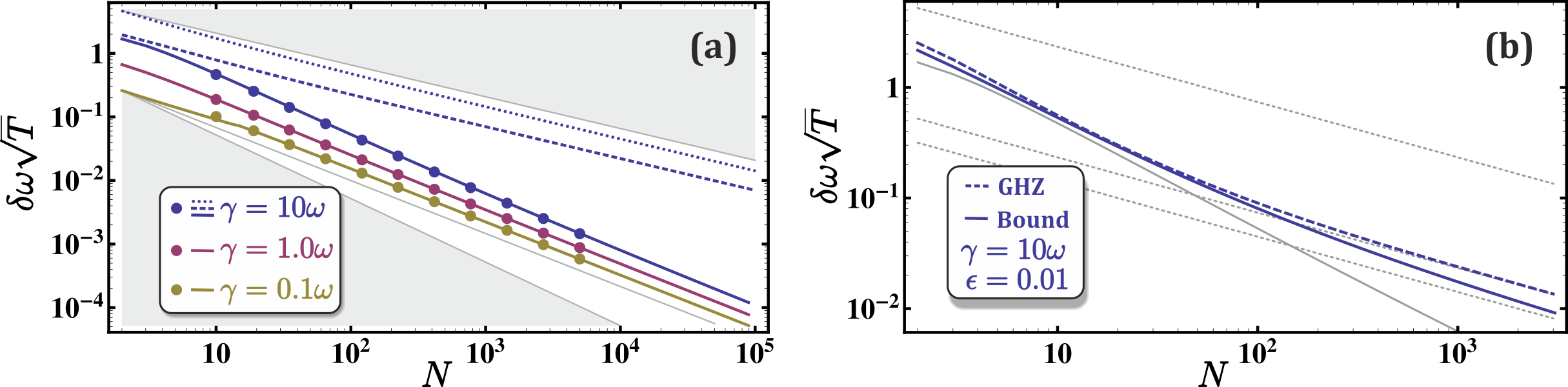}
\caption{\textbf{(a)} Precision scaling for bounds and GHZ strategy. The bound for parallel noise with optimisation of the single-round duration is shown (dotted), as well as the bounds for transversal noise without (dashed) and with (solid) optimisation. Without optimisation (taking $t$ optimal for a single qubit) the asymptotic scaling is still SQL-like, but with optimisation super-classical scaling is maintained. In the latter case, the bound is saturated by the GHZ-state strategy (dots). The thin line shows $1/N^{5/6}$ scaling for reference, and the borders of the upper and lower shaded regions show SQL-like and Heisenberg scaling respectively.  \textbf{(b)} Bound (solid) and GHZ-state strategy (dashed) for noise with a dominant transversal and a small parallel component, $\alpha_x/\alpha_z=99$. The scaling is super-classical for moderate $N$, while the $z$-component determines the asymptotic scaling which is hence SQL-like. The CE bound approaches $\sqrt{2\gamma\epsilon/N}$ (lower thin dotted) and the GHZ-state $\sqrt{2\gamma\epsilon e/N}$ (middle thin dotted). The finite-$N$ CE bound for parallel noise with the same strength is shown for reference (upper thin dotted) as well as for transversal noise (thin solid). In both \textbf{(a)} and \textbf{(b)} results are given in units where $\omega=1$.}
\label{fig.boundsGHZ}
\end{figure*}

%Noisy metrology beyond the SQL

\textit{Parallel noise---}
To understand the role of noise we compare parallel and perpendicular noise, as well as noise which is directional but not fully concentrated on any of the axes. We start by parallel dephasing, which has been studied before~\cite{huelga1997}.

%Dephasing alligned with the evolution

For a single qubit, the QFI for an optimal input state is given by $\mathcal{F}_{1}^{\text{opt}} = e^{-2t\gamma} t^2$. In the noiseless case ($\gamma\!=\!0$), as expected, the longer the probe state is allowed to evolve, the more information can be extracted about the parameter. In the noisy case, after a time $1/2\gamma$ the dephasing process wins over the unitary evolution and the extractable information is degraded. For a classical strategy using $N$ independent qubits, this is the optimal time and the SQL for parallel noise is
\begin{equation}
\label{ghz_cl_Zbounds}
\delta\omega\,\sqrt{T} \geq \sqrt{\frac{2 \gamma e}{N}}.
\end{equation}

It has been proven~\cite{escher2011} that when quantum strategies are allowed, the precision is instead bounded by
\begin{equation}
\label{dephZtbounds}
\delta \omega\,\sqrt{T} \geq
\sqrt{\frac{1+(e^{2\gamma t}-1)N}{t\,N^2}}
\geq
\sqrt{\frac{c_z(\gamma,t)}{N}} .
\end{equation}
The tighter bound coincides with the finite-$N$ CE method~\cite{kolodynski2012} for this channel and asymptotically reduces to the weaker bound. Optimising, we find $t_z^{\text{opt}}\!=\!w[N]/2\gamma$, where $w[N]\!=\!1+W\!\left[\frac{1-N}{eN}\right]$ and $W[z]$ is the Lambert W function. Asymtotically, $t_z^{\text{opt}}$ approaches zero as $1/\sqrt{N}$. Hence the asymptotic scaling of \eqnref{dephZtbounds} is dictated by the small-time behaviour of $c_z(\gamma,t)$, which is fully determined by the CS method \cite{demkowicz2012}, i.e. by the geometrical location of the channel in the convex set of all quantum qubit maps (see \figref{fig.CS}):
\begin{equation}
\label{eq.czexpansion}
c_z(\gamma,t) = \epsilon_z(t)^{2}t = 2 \gamma + 2\gamma^2 t + O(t^2).
\end{equation}
Using \eqnref{ghz_cl_Zbounds} and \eqnref{dephZtbounds} we see that the asymptotic scaling is SQL-like and that quantum strategies provide only a constant factor improvement of $\sqrt{e}$, as found earlier~\cite{huelga1997,escher2011}. Thus, for purely parallel dephasing, optimising $t$ leads only to a minor improvement of precision. The asymptotic scaling imposed by \eqnref{dephZtbounds} is shown in \figref{fig.boundsGHZ}(a).

The asymptotic SQL-like scaling is known to be saturable with spin squeezed states~\cite{ulam2001,kolodynski2012}. For comparison with transversal noise, we note that a GHZ input state gives no improvement over the SQL \eqnref{ghz_cl_Zbounds}. A GHZ strategy is thus useless for parallel noise.

%Dephasing perpendicular to the evolution

\textit{Transversal noise---}
We now turn our attention to perfectly transversal noise. While we do not have an analytical expression, we can efficiently compute the finite-$N$ CE bound and determine the optimal evolve-and-measure duration $t_x^{\text{opt}}$ for each $N$ numerically (see \figref{fig.boundsGHZ}(a)).

Without optimisation the asymptotic scaling is still restricted to SQL-like, as indeed it must be since transversal noise corresponds to a full rank channel ~\cite{demkowicz2012,fujiwara2008,ji2008,hayashi2011}. However, unlike for parallel noise, the asymptotic quantum improvement factor is not bounded by the geometry of the set of qubit maps (see \figref{fig.CS}). In consequence, at short times, the asymptotic SQL factor $c_x(\gamma,t)$ from the CS method is
\begin{equation}
\label{eq.cxexpansion}
c_x(\gamma,t) = \epsilon_x(t)^{2}t  = \frac{\gamma^2\omega^2}{12} t^3+ O(t^5) ,
\end{equation}
and, as opposed to \eqnref{eq.czexpansion}, $c_x(\gamma,t)\!\rightarrow\!0$, as $t\!\rightarrow\!0$.

Optimising the finite-$N$ CE bound over $t$ for each $N$, we find that super-classical scaling is maintained (verified numerically up to $N\!=\!10^8$ for $\gamma/\omega$ between 0.001 and 100) and follows an asymptotic behaviour very well described by
\begin{equation}
\label{eq.xboundscaling}
\delta\omega\,\sqrt{T} \ge \sqrt{ \frac{ c_x(\gamma)} {N^{5/3}} } .
\end{equation}
For large $N$, $t_x^{\text{opt}} \rightarrow 0$, since otherwise the restriction to SQL-like scaling applies. From our numerics we obtain that $t^{\text{opt}}_x = (3/\gamma\omega^2 N)^{1/3}$ and
\begin{equation}
\label{eq.xboundcoeff}
c_x(\gamma) = \frac{3^{2/3}}{2} (\gamma\omega^2)^{1/3} .
\end{equation}
Thus, at the level of bounds, the $t$-optimised quantum strategies provide a scaling rather than a constant factor improvement over classical schemes for transversal noise. To confirm this scaling, as we have not yet demonstrated that the bound is tight, we examine a specific strategy based on a GHZ state.

For given $N$, we can analytically compute the QFI corresponding to a GHZ input (see Appendix B). The expression becomes cumbersome with larger $N$, but we can numerically determine $t^{\text{opt}}_x$ and the minimum $\delta\omega$, for values of $N$ up to several thousands. The result is shown in \figref{fig.boundsGHZ}(a). Clearly, for the displayed values of $\gamma/\omega$, the GHZ state is optimal. What is more, the GHZ strategy shows no sign of returning to SQL-like scaling for large $N$ (verified for $N$ up to 5000 and $\gamma/\omega$ between 0.001 and 10). Note that $t_x^{\text{opt}}\!\rightarrow\!0$ with increasing $N$. If we expand the QFI for GHZ inputs to first order in $t$, we find that $\mathcal{F}_{N}/t = N^2t + O(t^2)$. As a semi-analytical check, substituting the numerically obtained $t^{\text{opt}}_x \!=\! (3/\gamma\omega^2 N)^{1/3}$, one recovers the scaling behaviour of the CE bound. Based on this strong numerical and semi-analytical evidence, we conjecture that the finite-$N$ CE bound is indeed tight for sufficiently large $N$, and that the asymptotic scaling is super-classical as predicted by \eqnref{eq.xboundscaling}.

%Intermediate noise

\textit{Intermediate noise---}
In a realistic implementation of a setup with transversal noise, most likely there will be deviations from perfect directionality. To account for such imperfections, in the following we consider deviations along the $z$ axis, such that $\alpha_{x}\!=\!1-\epsilon$ and $\alpha_z\!=\! \epsilon$ (we note however that similar conclusions hold for more general deviations, i.e.~taking also $\alpha_y \neq 0$ \cite{note}). Once there is some $z$-noise, the asymptotic scaling must return to SQL-like. If this were not the case, then by starting with a bit of $z$-noise and adding $x$-noise until $\alpha_x$ dominates, one could recover super-classical scaling. Hence, for large enough $N$, the precision would be improved by adding noise, which is clearly unphysical. Nevertheless, super-linear scaling can still persist over a large region of $N$.

Following this reasoning, the asymptotic bound must be of the form
\begin{equation}
\label{eq.xzbound}
\delta\omega\,\sqrt{T} \ge \sqrt{ \frac{c_{xz}(\gamma,\epsilon)}{N} }
\end{equation}
with $c_{xz}(\gamma,\epsilon) \geq c_z(\epsilon\gamma,0) \! =\! 2\epsilon\gamma$. In fact, we expect that equality must hold. This is because $t^\t{opt}_{xz}\!\rightarrow\!0$ for asymptotic $N$, but for very short times non-commutativity effects seize to apply and our model is equivalent to a process in which unitary evolution and noise along each axis are applied sequentially in any order. Equality is confirmed by the numerical results in \figref{fig.boundsGHZ}(b), where we see that the CE bound attains the asymptotic scaling $\sqrt{2\gamma\epsilon/N}$. The same argument applies to the GHZ-state strategy. In this case, we can see explicitly what happens at short times. Expanding up to second order in $t$ we have $\mathcal{F}_{N}/t = N^2t-\left[ (2N-1)(1-\epsilon)+4N^2\epsilon \right] N \gamma t^2 /2$. For large $N$, the expression reduces to the case of pure parallel noise with an effective noise strenght of $\epsilon \gamma$ and asymptotically the precision scales as $\sqrt{2\epsilon\gamma e/N}$. Thus, for $\epsilon\!>\!0$ the GHZ state no longer saturates the bound. In \figref{fig.boundsGHZ}(b) we explore the transition from super-classical to asymptotic SQL-like scaling for both the bound and the GHZ state. Intuitively, we expect that the transition point $N_{x\!z}(\gamma,\epsilon)$ to SQL-like behaviour increases smoothly to infinity for $\epsilon\!\rightarrow\!0$. Indeed, from \eqnref{eq.xboundscaling}, \eqnref{eq.xboundcoeff} and \eqnref{eq.xzbound} we can estimate $N_{xz}$ as the intersection of the super-classical and asymptotic SQL-like asymptotes, $N_{xz}(\gamma,\epsilon) \sim 3\omega/(8 \gamma \epsilon^{3/2})$. Beyond $N_{x\!z}(\gamma,\epsilon)$, no significant gain is obtained by increasing initial entanglement, i.e.~increasing the size of the entangled probe or using copies of disentangled probes of the same size leads to the same improvement in precision.

Finally, we remark that while $t^\text{opt}$ tends to zero for large $N$, in practice $t$ may be bounded from below by a finite resolution $t^\text{min}$. In this case, one can optimise $t$ for $N$'s up to the point where $t^\text{opt}\!=\! t^\text{min}$ and obtain the $1/N^{5/6}$ scaling of \figref{fig.boundsGHZ}(a) in this region. Beyond this point, one sets $t\!=\!t^\text{min}$, and the precision becomes restricted to SQL-like. Thus, the effect of $t^\text{min}\!>\!0$ for transversal noise, despite its different nature, has similar consequences to the effect of $\epsilon\!>\! 0$. Both lead to non-zero asymptotic constant factors, $c_{x}(\gamma,t)$ and $c_{xz}(\gamma,\epsilon)$ respectively, which can approach zero when $t^\text{min} \rightarrow 0$ or $\epsilon \rightarrow 0$. More generally, when both $\epsilon\!>\!0$ and $t^\t{min}\!>\! 0$, the ultimate precision is limited by the bound \eqnref{dephZtbounds} with $c_z(\gamma\epsilon,t^\t{min})$.%, which also vanishes for vanishing $\epsilon$ and $t^\t{min}$.

%Conclusions

\textit{Conclusions---}
Although recent results have shown that realistic noise prevents quantum metrology strategies from outperforming their classical counterparts by more than a constant factor, when the noise is independent of the probe size, it is nevertheless possible to observe scaling beyond the standard quantum limit in the presence of uncorrelated noise. Adapting the classical simulation and finite-$N$ channel extension methods~\cite{demkowicz2012,kolodynski2012}, we have shown that optimising the duration of evolve-and-measure rounds significantly enhances the precision. In particular, we have considered a unitary evolution with a well defined direction, and noise with a preferential direction transversal to the unitary. In this setting we have analysed a frequency estimation protocol and showed that the GHZ state achieves maximal precision in the presence of noise, providing a precision scaling of $1/N^{5/6}$. Furthermore, we have demonstrated that although the asymptotic scaling returns to SQL-like when the noise deviates from being perfectly transversal, the constant factor improvement of the precision can be significantly enhanced for small deviations.

We believe, our work opens an avenue towards useful quantum metrology protocols in realistic, noisy settings. In particular, we expect our model to capture the essential features of anisotropic noise occuring e.g. in quantum magnetometry~\cite{wasilewski2010,wolfgramm2010}. Our results indicate a gap in scaling between the SQL and the attainable precision when the geometry of the noise is accounted for, and we hope to stimulate further research as there is ample room for particular measurement schemes to achieve precisions inside this gap.

\begin{acknowledgements}
We would like to thank R. Demkowicz-Dobrza\'{n}ski, M. Horodecki,
P. Horodecki and R. Horodecki for helpful discussions.
This work was supported by the the EU Q-Essence project, the ERC
Starting Grant PERCENT, the Spanish FIS2010-14830 project, the
Foundation for Polish Science TEAM project and International PhD
project ``Physics of future quantum-based information
technologies'' (grant MPD/2009-3/4), the ERC grant QOLAPS, the NCN grant No. 2012/05/E/ST2/02352,
the ERA-NET CHIST-ERA project QUASAR and the Excellence Initiative of
the German Federal and State Governments (grant ZUK 43). We also thank B.~M.~Escher for pointing out
typos leading to erroneous units in several expressions, which have been corrected in this updated version (the uncorrected expressions corresponded to measuring time and frequency in units of $\omega$).
\end{acknowledgements}

\appendix

\section{Solving the master equation}

We write the map corresponding to evolution under the master equation (1) in the main text for time $t$ as a composite map of the form $\mathcal{E}_{\omega}^{\otimes N}$. Following Andersson \textit{et al.}~\cite{andersson2007}, the single-qubit maps are then given by
\begin{equation}
\mathcal{E}_{\omega} \left( \rho \right) = \sum_{i,j} S_{ij} \tilde{\sigma}_{i} \rho \tilde{\sigma}_{j} ,
\label{map}
\end{equation}
where the $\tilde{\sigma}_{i}$ are normalised variants of the Pauli operators $\tilde{\sigma}_{i} = \sigma_{i}/\sqrt{2}$, and all elements of the matrix S are zero, except $S_{00} =a+b$, $S_{11} =d+f$, $S_{22} =d-f$, $S_{33} =a-b$, $S_{03} =\imath c$, $S_{03} =-\imath c$. Note that in accordance with the main text, we denote the Pauli operators, so that ${\sigma}_{1}\!\equiv\!{\sigma}_{x}$, ${\sigma}_{2}\!\equiv\!{\sigma}_{y}$ and ${\sigma}_{3}\!\equiv\!{\sigma}_{z}$. For the application of known bounds for parameter estimation, the map $\mathcal{E}_{\omega}$ can easily be put on Kraus form~\cite{andersson2007} (see below). The coefficients $a,b,c,d$, and $f$ are real and depend on $\omega$, $\gamma$, and $t$. They are given by
\begin{widetext}
\begin{align}
a  & =\frac{1}{2}e^{-\frac{1}{2}t\left(  1+\alpha_{x}+\alpha_{y}-\alpha
_{z}\right)  \gamma}\left(  1+e^{t\left(  \alpha_{x}+\alpha_{y}\right)
\gamma}\right)  \nonumber \\
b  & =\frac{1}{2}e^{-\frac{1}{2}t\left(  \gamma+\alpha_{z}\gamma+\sqrt{\left(
\alpha_{x}-\alpha_{y}\right)  ^{2}\gamma^{2}-4\omega^{2}}\right)  }\left(
1+e^{t\sqrt{\left(  \alpha_{x}-\alpha_{y}\right)  ^{2}\gamma^{2}-4\omega^{2}}%
}\right)  \nonumber\\
d  & =\frac{1}{2}e^{-\frac{1}{2}t\left(  1+\alpha_{x}+\alpha_{y}-\alpha
_{z}\right)  \gamma}\left(  -1+e^{t\left(  \alpha_{x}+\alpha_{y}\right)
\gamma}\right)  \\
f  & =\frac{e^{-\frac{1}{2}t\left(  \gamma+\alpha_{z}\gamma+\sqrt{\left(
\alpha_{x}-\alpha_{y}\right)  ^{2}\gamma^{2}-4\omega^{2}}\right)  }\left(
-1+e^{t\sqrt{\left(  \alpha_{x}-\alpha_{y}\right)  ^{2}\gamma^{2}-4\omega^{2}}%
}\right)  \left(  \alpha_{x}-\alpha_{y}\right)  \gamma}{2\sqrt{\left(
\alpha_{x}-\alpha_{y}\right)  ^{2}\gamma^{2}-4\omega^{2}}}\nonumber\\
c  & =\frac{e^{-\frac{1}{2}t\left(  \gamma+\alpha_{z}\gamma+\sqrt{\left(
\alpha_{x}-\alpha_{y}\right)  ^{2}\gamma^{2}-4\omega^{2}}\right)  }\left(
-1+e^{t\sqrt{\left(  \alpha_{x}-\alpha_{y}\right)  ^{2}\gamma^{2}-4\omega^{2}}%
}\right)  \omega}{\sqrt{\left(  \alpha_{x}-\alpha_{y}\right)  ^{2}\gamma
^{2}-4\omega^{2}}}\nonumber
\end{align}
\end{widetext}

\section{GHZ input}

From the symmetry of $\ket{GHZ}$ and of the channel under permutation of the parties, one finds that the evolved state $\rho_{\omega}$ is diagonal in the GHZ basis, that is the basis formed by all states of the form  $\left( \ket{m_{1} \cdots m_{N}} \pm \ket{\overline{m}_1 \cdots \overline{m}_{N}}  \right)/\sqrt{2}$ with $m_{i}=0,1$ and $\ket{\overline{m}_i}=\sigma_1 \ket{m_i}$. We can then parameterize each density matrix element $\bra{m_i\ldots m_N}\rho_\phi\ket{m_i'\ldots m_N'}$ by the number $m=\sum m_{i}$. It is possible to show that the diagonal terms are given by
\begin{equation}
\left\langle m\right\vert \rho_{\phi}\left\vert m\right\rangle =\frac{1}{2}\left[ d^{m}a^{N-m} + d^{N-m}a^{m} \right] ,
\end{equation}
while the anti-diagonal terms fulfill $m^{\prime}=\sum m_{i}^{\prime}=N-m$, and hence
\begin{equation}
\left\langle m\right\vert \rho_{\phi}\left\vert m^{\prime}\right\rangle =\frac{1}{2}\left[ f^{m}\left(b-\imath c\right)^{N-m} + f^{N-m}\left(b+\imath c\right)^{m} \right] .
\end{equation}
Each $m$-term has a degeneracy of $\binom{N}{m}$. Using this parametrisation, the diagonalization of the $2^N \times 2^N$ density matrix is reduced to the diagonalization of $\left\lfloor N/2 +1\right\rfloor$ different $2 \times2 $ density matrices ($\lfloor \cdot \rfloor$ denotes the floor function) (see for instance \cite{ChavesR2012}). \\

\section{Details on the channel extension method for finite $N$}
\label{deviations}

The channel extension (CE) method~\cite{demkowicz2012,kolodynski2012} uses the fact that allowing the channel to act trivially on an extended space can only increase the precision of estimation. Hence, for $N$ independent channels acting on an input state $\rho_{in}^N$ of $N$ qubits
\begin{equation}
\label{eq.ceboundstart}
\max_{\rho_\t{in}^N}\mathcal{F}_N\!\left[\mathcal{E}_{\omega}^{\otimes N}\!\left[\rho_\t{in}^N\right]\right]
\le
\max_{\rho_\t{in}^{2N}}\mathcal{F}_N\!\left[\left(\mathcal{E}_{\omega}\!\otimes\mathbb{I}\right)^{\otimes N}\!\left[\rho_\t{in}^{2N}\right]\right],
\end{equation}
As a result of this extension of the input state space from $N$ to $2N$ particles, another upper bound on the QFI can be obtained from \eqref{eq.ceboundstart}, which does not involve any input state optimisation~\cite{fujiwara2008}. For any input state
\begin{equation}
\label{CEbound}
\mathcal{F}_{N}(t)\leq4\,\min_{K} \left\{ N\left\Vert\alpha_{K}(t)\right\Vert +N(N-1)\left\Vert\beta_{K}(t)\right\Vert^2\right\}\! ,
\end{equation}
where we have made the dependence on the evolution time (which fixes the channels) explicit. The $\left\Vert \cdot\right\Vert$ denotes the operator norm and
\begin{equation}
\begin{split}
\alpha_{K}(t) & =  \sum_{i}\!\dot{K}_{i}^{\dagger}\!(t)\dot{K}_{i}(t) , \\
\beta_{K}(t) & =  \mathrm{i}\sum_{i}\!\dot{K}_{i}^{\dagger}\!(t)K_{i}(t)
\end{split}
\end{equation}
with $\dot{K}_{i}(t)\!=\!\partial_{\omega}K_{i}(t)$. The minimization in \eqnref{CEbound} is performed over all locally equivalent Kraus representations of $\mathcal{E}_{\omega}$~\cite{demkowicz2012}. These are generated by all $r\!\times\!r$ Hermitian matrices $\mathbf{h}(t)$, so that $\tilde{\mathbf{K}}(t)\!=\!\mathbf{K}(t)$ and $\dot{\tilde{\mathbf{K}}}(t) \!=\! \dot{\mathbf{K}}(t) \!-\! \mathrm{i} \mathbf{h}(t) \mathbf{K}(t)$, where $\mathbf{K}(t)$ is a column vector containing any starting $\left\{\!K_i(t)\right\}_{i=1}^r$ as its elements and $[\mathbf{h}\,\mathbf{K}]_i\!=\!\sum_j h_{ij}K_j$. We construct matrices
\begin{equation}
\begin{split}
\mathbf{A}(t) & =
\left[\begin{array}{cc}
\sqrt{\lambda_{a}}\,\openone_{2} & \dot{\mathbf{K}}(t)^{\dagger}\\
\dot{\mathbf{K}}(t) & \sqrt{\lambda_{a}}\,\openone_{2r}
\end{array}\right]\! , \\
\mathbf{B}(t) & =
\left[\begin{array}{cc}
\sqrt{\lambda_{b}}\,\openone_{2} & \left(\mathrm{i}\dot{\mathbf{K}}(t)^{\dagger}\,\mathbf{K}(t)\right)^{\dagger}\\
\mathrm{i}\dot{\mathbf{K}}(t)^{\dagger}\,\mathbf{K}(t) & \sqrt{\lambda_{b}}\,\openone_{2}
\end{array}\right]\!,
\end{split}
\end{equation}
with $\openone_d$ representing a $d\!\times\!d$ identity matrix, and $\lambda_{a/b}$ being some real parameters. One can then see that the minimisation problem of \eqnref{CEbound} is equivalent to a semi-definite programming task. This is because the positive semi-definiteness of $\mathbf{A}(t)$ and $\mathbf{B}(t)$ corresponds respectively to the conditions
\begin{equation}
\begin{split}
\alpha_{K}(t)  & = \dot{\mathbf{K}}^{\dagger}\!(t)\,\dot{\mathbf{K}}(t) \leq \lambda_a \openone_{2} , \\
\beta_{K}^\dagger\!(t)\beta_{K}(t) & = \mathbf{K}^{\dagger}\!(t)\dot{\mathbf{K}}(t)\dot{\mathbf{K}}^\dagger\!(t)\mathbf{K}(t) \leq \lambda_b \openone_{2} ,
\end{split}
\end{equation}
so that \eqnref{CEbound} can be rewritten as
\begin{widetext}
\begin{equation}
\label{CEboundSDP}
\mathcal{F}_{N}(t) \le 4\min_{\mathbf{h}(t)}\left\{ N\,\lambda_{a}+N(N-1)\lambda_{b}\right\} \hspace{1cm} \textrm{subject to: }\mathbf{A}(t)\ge0,\,\mathbf{B}(t)\ge0 .
\end{equation}
\end{widetext}
This construction has been introduced in~\cite{kolodynski2012} as the finite-$N$ CE method, proving that an upper bound on $\mathcal{F}_N(t)$ can be efficiently computed for any $N$. Its asymptotic version, which leads to a weaker bound, has been described in~\cite{demkowicz2012} and corresponds to adding a further constraint on $\mathbf{h}(t)$, namely that $\beta_K(t)\!=\!0$, so that $\lambda_b\!=\!0$ and the second term in \eqref{CEboundSDP} vanishes. However, in our present analysis we need to be more careful, as the $t$-optimisation results in $t$ depending $N$. For fixed $t$ the asymptotic method of Ref.~\cite{demkowicz2012} directly applies, so that, taking $\beta_K(t)\!=\!0$, from the first term of \eqnref{CEbound} we obtain the asymptotic constant factor improvements for our model.

To allow for optimisation over the evolve-and-measure round duration $t$, we adapt \eqnref{CEboundSDP} and maximise the bound
\begin{equation}
\label{CEboundFdivt}
\frac{\mathcal{F}_{N}(t)}{t} \le 4\left[ N\frac{\lambda_{a}^{\t{min}}(t)}{t}+N(N\!-\!1)\frac{\lambda_{b}^\t{min}(t)}{t}\right]\!,\\
\end{equation}
over $t$, where $\lambda_{a/b}^\t{min}(t)$ correspond to the optimal $\lambda_{a/b}$ in \eqnref{CEboundSDP} obtained after performing the SDP minimization over $\mathbf{h}(t)$ for given $t$. After finding the optimal $t\!=\!t^\t{opt}(N)$ for each $N$ in \eqnref{CEboundFdivt}, we arrive at the bound applied in the paper.

Note that, as the second term in \eqnref{CEboundFdivt} depends on $N$ through both $\lambda_b^\t{min}$ and $t^\t{opt}$, it is not true that in the limit $N\!\rightarrow\!\infty$ it is optimal to set  $\lambda_b\!=\!0$ as in the asymptotic CE method~\cite{demkowicz2012}. For transversal noise, based on numerical results we determine the short-times asymptotic SQL constant factor to be $c_x(\gamma,t)\!=\!\frac{\gamma\omega^2}{6}t^2\!+\!O(t^3)$ which improves the analytically computed geometrical bound of eq.~(8) in the main text. However, if one substitutes in this formula the $t_{\t{opt}}(N)$ from the finite-$N$ CE method (given above eq.~(10) in the main text), one would arrive at a non-saturable asymptotic bound below the correct one (given by eq.~(10)) by a factor of 3. This fact emphasizes the applicability of the finite-$N$ CE method.

For the purpose of computing the $t$-optimised, finite-$N$ CE bound we have implemented a semi-definite program using the CVX package for Matlab~\cite{CVX}.


\begin{thebibliography}{31}
\bibitem [{not()}]{note}%
  \BibitemOpen
  \href@noop {} {}\bibinfo {note} {We find (analytically for the GHZ state and
  numerically for the finite-$N$ CE bound) that when any two noise components
  are non-zero, asymptotic scaling returns to SQL-like. In particular, having
  two transversal components present simultaneously is enough. However, as
  discussed in the main text, it is the parallel noise that, if present,
  dictates the asymptotic behaviour and is less likely to be eliminated in
  realistic experimental scenarios.}\BibitemShut {Stop}%
\end{thebibliography}
\end{document}